\documentclass[a4paper,runningheads]{llncs}

\usepackage[british]{babel}

\usepackage{mathrsfs}
\usepackage{amsmath}
\usepackage{amsfonts}
\usepackage{amssymb}
\usepackage{mathpartir}
\usepackage[only,fatsemi]{stmaryrd}
\usepackage{tikz}
\usetikzlibrary{arrows.meta,calc}
\usepackage{hyperref}
\usepackage[capitalise,noabbrev]{cleveref}
\usepackage{adjustbox}
\usepackage{orcidlink}
\usepackage{cite} %

\usepackage{mathrsfs}
\usepackage{listings}
\usepackage[T1]{fontenc}
\usepackage{graphicx}
\usepackage{amsmath}
\usepackage{amsfonts}
\usepackage{amssymb}
\usepackage{mathpartir}
\usepackage[only,fatsemi]{stmaryrd}
\usepackage{upgreek}
\usepackage{wrapfig}
\usepackage{tikz}
\usetikzlibrary{arrows.meta,calc,decorations.markings,shapes.misc}

\tikzset{
  >=Stealth,
  lab/.style={midway,fill=white,inner sep=2pt},
  markarrows/.style={
    postaction={decorate},
    decoration={markings,
      mark=at position .85 with {\arrow{Stealth}},
      mark=at position 1 with {\arrow{Stealth}}}
  }
}

\newcommand{\rom}[1]{\uppercase\expandafter{\romannumeral #1\relax}}

\newcommand{\metaDef}{\mathrel{\mathop:}=}
\newcommand{\tlint}[1]{\tau@#1}
\newcommand{\tlthen}[1]{#1\text{.then}}
\newcommand{\tlelse}[1]{#1\text{.else}}
\newcommand{\gencom}{p.e \rightarrow q.x}
\newcommand{\gensel}{p \rightarrow q[L]}
\newcommand{\pn}{\mathrm{pn}}
\newcommand{\llbracket}{[\![}
\newcommand{\rrbracket}{]\!]}
\newcommand{\MCL}{\mathscr{L}}

\usepackage[
  disable,
  textwidth=\marginparwidth,
  colorinlistoftodos,
  prependcaption,
  linecolor=green,backgroundcolor=green!25,bordercolor=green
  ]{todonotes}
\makeatletter
\if@todonotes@disabled
\else
\let\tmptitle\title
\renewcommand{\title}[1]{\tmptitle{#1\smash{\raisebox{.2em}{\hspace{.7em}\makebox[0pt][c]{\scalebox{0.4}{\normalfont\parbox[c]{3.2em}{\centering\colorbox{red!20}{\parbox{3.2em}{\centering todos\\enabled}}}}}}}}}
\fi
\makeatother

\begin{document}

\title{Type-Based Enforcement of Non-Interference for Choreographic Programming}

\titlerunning{Type-Based Secure Choreographies}

\author{Marco Bertoni \inst{1}\orcidlink{0009-0000-9224-6005} \and
Saverio Giallorenzo \inst{2,3}\orcidlink{0000-0002-3658-6395} \and
Marco Peressotti \inst{4}\orcidlink{0000-0002-0243-0480}}

\authorrunning{M. Bertoni et al.}

\institute{Inria Paris, France \and Università di Bologna, Italy \and 
Inria Sophia Antipolis, France \and FORM, University of Southern Denmark, Denmark}
\maketitle

\begin{abstract}
Choreographies describe distributed protocols from a global viewpoint, enabling
correct-by-construction synthesis of local behaviours. We develop a
policy-parametric type system that prevents information leaks from high-security
data to $low$-security observers, handling both explicit and implicit flows
through a program-counter discipline. The system supports recursive procedures
via a procedure context that we reconstruct through constraint generation. We
prove termination-insensitive non-interference with respect to a standard
small-step semantics.
\end{abstract}

\section{Introduction}
\label{sec:intro}

Distributed software implements protocols through communicating components --
authenticating users, reconciling payments, etc.. Two paramount
concerns in distributed software are \emph{correct coordination} among
participants and \emph{confidentiality} of handled data. Choreographic
programming languages~\cite{montesi2023introduction} address coordination by
describing protocols from a global viewpoint, from which one can derive
compliant distributed software. This work tackles the second concern:
\emph{ensuring that choreographies do not leak secret information to public
observers}.

The challenges of ensuring confidentiality emerge in realistic protocols where
control flow depends on sensitive data. 
We exemplify the concept with a simple password-recovery scenario. A requester
$\mathtt{r}$ submits an email address to a service $\mathtt{s}$, which may then
task a mailer service $\mathtt{m}$ to send a reset link. While the protocol
concerns a few straightforward steps, even seemingly innocuous design choices
can inadvertently reveal sensitive information. Whether the email corresponds to
a registered account is sensitive information: revealing it enables username
enumeration attacks. Thus, we must treat the outcome of a check
$\mathtt{exists(email)}$ by $\mathtt{s}$ as secret, while the behaviour
observable by $\mathtt{r}$ is considered public.

Let us visualise the issue with an insecure choreography for that scenario. In
the choreography, \texttt{a.e -> b.x} means that \texttt{a} sends the evaluation
of expression \texttt{e} to \texttt{b}, which stores the received value into its
local variable \texttt{x}.

\medskip
\noindent\begin{minipage}{.18\textwidth}
{\begin{center}
  \vspace{.5em}\color{gray}Insecure\\ choreography\\ example
\end{center}}
\end{minipage}
\vrule\;
\begin{minipage}{.7\textwidth}
\begin{verbatim}
r.email -> s.email
if s.exists(email) then
  s.email -> m.email; s."email sent" -> r.msg
else s."unknown user" -> r.msg
\end{verbatim}
\end{minipage}
\medskip

Although the boolean returned by $\mathtt{exists(email)}$ is never sent directly
to $\mathtt{r}$, the requester can distinguish executions by observing different
response messages. As a consequence, $\mathtt{r}$ learns the sensitive bit of
information: whether the account exists. The leak arises not from an explicit
transmission of secret data, but from a control-flow dependency that produces
observable differences.

We can remove the leak by ensuring that branching on $\mathtt{exists(email)}$
affects only internal communication with the mailer, while keeping the
observable behaviour at $\mathtt{r}$ uniform.

\medskip
\noindent\begin{minipage}{.18\textwidth}
{\begin{center}
  \vspace{.5em}\color{gray}Secure\\ choreography\\ variant
\end{center}}
\end{minipage}
\vrule\;
\begin{minipage}{.7\textwidth}
\begin{verbatim}
r.email -> s.email
if s.exists(email) then s.email -> m.email else skip
s."check your inbox" -> r.msg
\end{verbatim}
\end{minipage}
\medskip

\noindent In the secure variant, executions that differ only in the secret predicate produce identical observations at $\mathtt{r}$. 
This example illustrates \emph{implicit flow} of information, where secrets influence which actions occur, in contrast to \emph{explicit flow}, where secret values are directly assigned or transmitted to public locations.

To detect and prevent such leaks systematically, we require a formal account of
how information may flow (both explicitly and implicitly) in a choreography. To
this end, following Denning~\cite{denning1976lattice}, we assign each datum a
\emph{security class} drawn from a lattice, whose order relation
$\sqsubseteq$ models relative sensitivity. Together with the class assignment,
the lattice structure defines a \emph{flow policy}, specifying when information
is allowed to move between locations of different sensitivity: data labelled
$\ell_1$ may flow to contexts labelled $\ell_2$ only if $\ell_1 \sqsubseteq
\ell_2$.

In this work, we focus on \emph{non-interference}~\cite{goguen1982security},
which requires that variations in higher-security inputs must not influence
lower-security outputs~\cite{volpano1996sound}. Intuitively, secure data should
not affect what an external observer can learn from a program's execution, i.e.,
interfere with the observed behaviour -- precisely the violation exhibited by
our first attempt at implementing password recovery.

We deem that tackling non-interference for choreographic programming offers
several advantages. First, all communication between processes appears
explicitly in the choreography, making potential information-flow channels
syntactically visible %
 -- in our example, the message from $\mathtt{s}$ to $\mathtt{r}$ immediately
identifies a possible leak. Second, security is verified for the entire protocol
through a single global judgement, rather than by composing separate guarantees
for individual participants. Third, choreographies capture both local control
flow and inter-process coordination, allowing us to detect implicit flows that
arise when a choice in one participant influences the observable behaviour of
others.
Finally, once a choreography is shown to satisfy non-interference, endpoint
projection preserves this property by construction~\cite{CMP23}. Thus, we obtain
end-to-end guarantees, from a verified global specification to the generated
distributed implementation. This combination of global reasoning, explicit
communication structure, and correctness-preserving compilation makes
choreographic programming a natural foundation for enforcing non-interference in
distributed systems.

\paragraph{Our approach.}
We develop a \emph{type-based enforcement} of non-interference for the reference language for procedural choreographic programming \cite{montesi2023introduction}.
Our type system is \emph{policy-parametric}: given a user-specified flow
policy over a lattice $\MCL$ and an observation level $low$, the type system
rejects choreographies where high-security data could influence $low$-observable
behaviour. 

Our type system centres on a judgement of the form
\(
\Delta;\Gamma;pc \vdash C
\),
where $\Gamma$ assigns security classes to variables, $pc \in \MCL$ tracks the current control level (providing the context for capturing implicit flow), and $\Delta$ specifies how procedures may be safely invoked at different control levels. 
This definition generalises standard information-flow type systems to the choreographic setting, where information flows both through message-passing between processes and through control dependencies in the global protocol.

We rely on two fundamental technical devices to obtain our results, embodied by
$pc$ and $\Delta$. 
First, we adopt a standard \emph{program-counter discipline}: the control
level $pc$ records the influence of secret data on the current execution point. 
When branching on a secret (like $\mathtt{exists(email)}$ in our example),
$pc$ is raised, and subsequent actions are checked in this elevated
context. This mechanism prevents $low$-observable behaviour from depending on high-security control flow, ruling out the implicit flow in our first password-recovery example.
Second, to support modular verification in the presence of recursion,
we introduce a \emph{procedure context} $\Delta$ that assigns to each
procedure and control level a set of admissible flow policies.
Intuitively, $\Delta$ serves as a contract:
$\Delta;\Gamma;pc \vdash X(\bar p)$ ensures that calling the procedure $X(\bar p)$ under $\Gamma$ at control level $pc$ preserves non-interference.
We develop a constraint-based reconstruction procedure that computes
$\Delta$ from the procedure definitions themselves,
thereby automating this verification step. %

Our enforcement target is a standard small-step semantics for choreographies. 
Following established practice in language-based information-flow security~\cite{volpano1996sound}, we adopt a \emph{termination-insensitive} attacker model, in which divergence is not considered observable. 
While stronger notions exist, this choice strikes a pragmatic balance between expressiveness (it allows programmes whose runtime depends on secrets) and security guarantees.

Our main result (\cref{thm:main}) establishes the soundness of the
type system with respect to termination-insensitive non-interference.

\paragraph{Structure of the Paper.}

In \cref{sec:chor}, we introduce the reference choreographic language and its
small-step operational semantics. In \cref{sec:types}, we present a
policy-parametric information-flow type system for the reference choreographic
language, tracking both explicit and implicit flow and supporting recursive
procedures given a valid procedure context $\Delta$. In \cref{sec:soundness} we
define and prove a soundness theorem, establishing that well-typed
choreographies satisfy termination-insensitive non-interference w.r.t.\@ the
language semantics. In \cref{sec:delta}, we define and prove the correctness of
a constraint-based reconstruction algorithm that, given a collection of
procedure definitions, computes a context $\Delta$ under which these procedures
are well-typed -- the expected correctness criteria. We discuss related work and
concluding remarks in, respectively, \cref{sec:related_work,sec:conclusion}.

\section{Preliminaries on Choreographies}
\label{sec:chor}

We recall essential notions of choreographic programming
languages~\cite{montesi2023introduction} and introduce the language intended for
users to write distributed programs, called \emph{Recursive Choreographies}. Choreographies formally describe intended
collaborative behaviour of processes in concurrent and distributed systems,
acting as protocols specifying how components interact to achieve shared goals.
Although choreographies are written is a way that reminds traditional local
programs, they execute through transitions representing distributed
communication and computation steps. A choreography represents the global state
of all participants and encompasses the actions that collectively implement
the protocol.

\subsection{Syntax}

\emph{Processes} are participants in a choreography, performing local computation and communicating with others. We range over process names by $p, q, r, s \in \mathsf{PName}$.

We define Recursive Choreographies by the context-free grammar:
\def\desc#1{\quad\parbox{.57\textwidth}{\small #1}}
\[
\begin{array}{rrll}
\mathscr{C} & ::= & \{X_i(\bar{p}_i) = C_i\}_{i \in I} & \desc{choreographic procedure definitions}\\
C & ::= & I; C & \desc{instruction sequencing} \\
  & \mid & \boldsymbol{0} & \desc{terminated choreography} \\
I & ::= & p.e \rightarrow q.x & \desc{$p$ sends the result of $e$ to $q$ which assigns it to $x$} \\
  & \mid & p \rightarrow q[L] & \desc{$p$ sends the constant $L$ to $q$} \\
  & \mid & p.x := e & \desc{$p$ assigns to $x$ the result of $e$} \\
  & \mid & \text{if } p.e \text{ then } C_1 \text{ else } C_2 & \desc{$p$ evaluates $e$ and chooses to continue as $C_1$ or $C_2$} \\
  & \mid & X(\bar{p}) & \desc{call procedure $X$}\\
  & \mid & \color{gray} r:X(\bar{p}).C' & \desc{\color{gray} (runtime term, $r$ has yet to enter the call)}\\
e & ::= & v \mid x \mid f(\bar{e}) & \desc{local values, variables, and function calls}
\end{array}
\]

Here $\mathscr{C}$ denotes a context of procedure definitions; $C$ denotes a choreography -- either terminated $\boldsymbol{0}$ or that composes in sequence instruction $I$ and continuation C. $I$ denotes instructions; from top to bottom, we find communication, selection, local assignment, conditional, and procedure call. $e$ denotes local expressions (constants $v \in \mathsf{Val}$, variables $x \in \mathsf{Var}$, function calls $f(\bar{e})$).

\subsection{Semantics}

We define the semantics of Recursive Choreographies using a small-step operational semantics~\cite{plotkin2004origins}, forming a labelled transition system. Configurations have form $\langle C, \Sigma, \mathscr{C} \rangle$ where $C$ is the choreography, $\mathscr{C}$ the procedure context, and $\Sigma$ the \emph{choreographic store}.

A \emph{process store} $\sigma : \mathsf{Var} \to \mathsf{Val}$ models one process's memory. We call $\mathsf{PStore}$ for the set of all process stores. A \emph{choreographic store} $\Sigma : \mathsf{PName} \to \mathsf{PStore}$ models the entire system's memory. We write $\Sigma[p.x \mapsto v]$ for updating store $\Sigma$ so local variable $x$ of process $p$ maps to $v$. We call $\mathsf{CStore}$ the set of all choreographic stores.

Given process store $\sigma$, expression $e$, and value $v$, we write $\sigma
\vdash e \downarrow v$ when $e$ evaluates to $v$ under $\sigma$. We do not
specify a system for deriving propositions of the kind $\vdash f(\bar{v})
\downarrow v$, since it is not important for our development: this system would
depend on how functions are defined, which we choose to abstract from. Instead,
we just assume that such a system exists, and that for any $f$ and
$\bar{v}$, it is always possible to derive $\vdash f(\bar{v})$ for some
$\bar{v}$.

Transition labels record the kind of step performed -- internal at $p$ ($\tlint{p}$), communication $p.v \rightarrow q$, selection ($\gensel$), or the outcome of a conditional at $p$ ($\tlthen{p}$/$\tlelse{p}$).

\begin{figure}[t]
\begin{adjustbox}{max width=\textwidth}
\begin{minipage}{1.043\textwidth}
\begin{mathpar}
	\inferrule*[lab=\textsc{local}] { \Sigma(p) \vdash e \downarrow v }
    {\langle p.x \metaDef e ; C, \Sigma, \mathscr{C} \rangle \xrightarrow{\tlint{p}} \langle C, \Sigma[p.x \mapsto v], \mathscr{C} \rangle}
	\and
	\inferrule*[lab=\textsc{com}] { \Sigma(p) \vdash e \downarrow v } 
    {\langle \gencom ; C, \Sigma, \mathscr{C} \rangle \xrightarrow{p.v \rightarrow q} \langle C, \Sigma[q.x \mapsto v], \mathscr{C} \rangle}
	\and
	\inferrule*[lab=\textsc{sel}] {  } 
    {\langle \gensel ; C, \Sigma, \mathscr{C} \rangle \xrightarrow{\gensel} \langle C, \Sigma, \mathscr{C} \rangle}
	\and
	\inferrule*[lab=\textsc{cond-then}] { \Sigma(p) \vdash e \downarrow \mathit{true}} 
    {\langle \text{if } p.e\text{ then } C_1\text{ else } C_2 ; C, \Sigma, \mathscr{C} \rangle \xrightarrow{\tlthen{p}} \langle C_1 \fatsemi C, \Sigma, \mathscr{C} \rangle }
	\and
	\inferrule*[lab=\textsc{cond-else}] { \Sigma(p) \vdash e \downarrow v \\ v \neq \mathit{true}} 
    {\langle \text{if } p.e\text{ then } C_1\text{ else } C_2 ; C, \Sigma, \mathscr{C} \rangle \xrightarrow{\tlelse{p}} \langle C_2 \fatsemi C, \Sigma, \mathscr{C} \rangle }
	\and
	\inferrule*[lab=\textsc{call-first}] { X(\bar{q}) = C \in \mathscr{C} \\
  \{p_0, \ldots, p_n\} = \{ r_0, r_1, \ldots, r_n\}}
	{\langle X(\bar{p});C', \Sigma, \mathscr{C} \rangle \xrightarrow{\tlint{r_0}} \langle r_1 : X(\bar{p}).C'; \ldots; r_{n} : X(\bar{p}).C'; C[\bar{q}/\bar{p}] \fatsemi C', \Sigma, \mathscr{C} \rangle}
	\and
	\inferrule*[lab=\textsc{call-enter}] {  }
	{\langle q : X(\bar{p}).C'; C, \Sigma, \mathscr{C} \rangle \xrightarrow{\tlint{q}} \langle C, \Sigma, \mathscr{C} \rangle}
	\and
	\inferrule*[lab=\textsc{delay}] {\langle C, \Sigma, \mathscr{C}\rangle \xrightarrow{\mu} \langle C', \Sigma', \mathscr{C}\rangle \\ \pn(I)\cap\pn(\mu) = \emptyset} 
    {\langle I; C, \Sigma, \mathscr{C}\rangle \xrightarrow{\mu} \langle I; C', \Sigma', \mathscr{C}\rangle}
	\and
	\inferrule*[lab=\textsc{delay-cond}] {\langle C_1, \Sigma, \mathscr{C}\rangle \xrightarrow{\mu} \langle C_1', \Sigma', \mathscr{C}\rangle \\ \langle C_2, \Sigma, \mathscr{C}\rangle \xrightarrow{\mu} \langle C_2', \Sigma', \mathscr{C}\rangle \\ p\notin \pn(\mu)}
    {\langle \text{if } p.e\text{ then } C_1\text{ else } C_2 ; C, \Sigma, \mathscr{C} \rangle \xrightarrow{\mu} \langle \text{if } p.e\text{ then } C_1'\text{ else } C_2' ; C, \Sigma', \mathscr{C} \rangle}
\end{mathpar}
\end{minipage}
\end{adjustbox}
\caption{Operational Semantics.}
\label{fig:semantics}
\end{figure}

We report the rules of the semantics of Recursive Choreographies in
\cref{fig:semantics}. Rule \textsc{com} models the communication of a value
computed at runtime: $p$ evaluates the expression $e$ against its local store
and sends the result $v$ to $q$ which stores it under variable $x$ in its local
store. Similarly, Rule \textsc{sel} models the communication of a constant $L$,
called a selection.\footnote{The construct of selections plays a critical role
in the handling of branching constructs in endpoint projection which is beyond
the scope of this work. We refer the curious reader to
\cite[Chapter~6]{montesi2023introduction}.} Rules \textsc{cond-then},
\textsc{cond-else} model conditional branching: process $p$ evaluates the guard
$e$ locally and the choreography continues as the corresponding branch -- the
operator $\fatsemi$ implements sequential composition ($C;C'$ is not part of the
grammar) by grafting the continuation $C$ in place of any $\boldsymbol{0}$ in
$C_1$ and $C_2$. Rules \textsc{call-first} and \textsc{call-enter} model
decentralised calls to a choreographic procedure i.e., without synchronising the
processes that enter the procedure. These rules rely on the runtime term
$r:X(\bar{p}).C'$ to track that process $r$ has not entered the call and thus
cannot perform any action from its continuation via the delay rules. Rule
\textsc{call-first} unfolds the procedure definition from the context
$\mathscr{C}$ and instantiates formal process parameters $\bar{q}$ with the
actual ones $\bar{p}$ in its body $C$. The selection of which process enters the
call first and which will do it later via \textsc{call-enter} is
nondeterministic since the premise $\{p_0, \ldots, p_n\} = \{ r_0, r_1, \ldots,
r_n\}$ allows to arbitrarily reindex $\bar{p}$. Rules \textsc{delay} and
\textsc{delay-cond} model that actions at independent processes are executed
concurrently by delaying the execution of an instruction or branch selection.

\section{Type System}
\label{sec:types}

We develop a type system inspired by non-interference standard techniques~\cite{myers2011proving,wright1994syntactic}, adapted to choreographies.
\emph{Security labels} are elements of a complete lattice $(\MCL, \sqsubseteq)$ with bottom $\bot$ satisfying $\bot \sqsubseteq l$ for every $l \in \MCL$. Labels capture Denning's security classes; we assign a security label to every variable.
A \emph{process security labelling} $\gamma : \mathsf{Var} \to \MCL$ assigns security labels to variables accessed by a process. A \emph{choreographic security labelling} $\Gamma : \mathsf{PName} \to (\mathsf{Var} \to \MCL)$ maps process names to their respective labellings.
The flow policy forbids information flow from an object with higher security label towards an object with lower associated label, thus defining confidentiality.

\begin{figure}[t]
\begin{mathpar}
\inferrule*[lab=\textsc{t-const}]{~}{\gamma \vdash v : \bot}
\and
\inferrule*[lab=\textsc{t-var}]{~}{\gamma \vdash x : \gamma\,x}
\and
\inferrule*[lab=\textsc{t-lfun}]{\gamma\vdash e_1:\ell_1 \;\; \cdots \;\; \gamma\vdash e_n:\ell_n \\ \ell' = \bigsqcup\nolimits_{i=1}^{n} \ell_i}{\gamma\vdash f(e_1,\dots,e_n):\ell'}
\and
\inferrule*[lab=\textsc{t-local}]{\Gamma\,p \vdash e : \ell' \quad \ell' \sqcup pc \;\sqsubseteq\; \Gamma\,p.x}{\Delta;\Gamma;pc \vdash \; p.x := e}
\and
\inferrule*[lab=\textsc{t-com}]{\Gamma\,p \vdash e : \ell' \quad \ell' \sqcup pc \;\sqsubseteq\; \Gamma\,q.x}{\Delta;\Gamma;pc \vdash \; p.e \to q.x}
\and
\inferrule*[lab=\textsc{t-sel}]{~}{\Delta;\Gamma;pc \vdash \; p \to q[L]}
\and
\inferrule*[lab=\textsc{t-cond}]{\Gamma\,p \vdash e : \ell' \quad \Delta;\Gamma;\,\ell' \sqcup pc \vdash C_1 \quad \Delta;\Gamma;\,\ell' \sqcup pc \vdash C_2}{\Delta;\Gamma;pc \vdash \; \mathbf{if}\;p.e\;\mathbf{then}\;C_1\;\mathbf{else}\;C_2}
\and
\inferrule*[lab=\textsc{t-proc}]{\Gamma' \in \Delta\, X\, pc \quad \Gamma[\bar{q}\mapsto \bar{p}] \equiv_{\{\bar{q}\}} \Gamma'}{\Delta;\Gamma;pc \vdash \; X(\bar{p})}
\and
\inferrule*[lab=\textsc{t-seq}]{\Delta;\Gamma;pc \vdash I \quad \Delta;\Gamma;pc \vdash C}{\Delta;\Gamma;pc \vdash \; I \,;\ C}
\and
\inferrule*[lab=\textsc{t-nil}]{~}{\Delta;\Gamma;pc \vdash \; \boldsymbol{0}}
\end{mathpar}
\caption{Typing Judgement.}
\label{fig:types}
\end{figure}

The type judgement is defined on the syntax of Recursive Choreographies as the smallest relations following the inference schemata presented in \cref{fig:types}.
Rules \textsc{t-const}, \textsc{t-var}, \textsc{t-lfun} type expressions $\Gamma\,p \vdash e : \ell$, where $\ell \in \MCL$ characterises the sensitivity of $e$.
Functions take the join of the labels of the arguments, thus assuming that local functions preserve labels and do not introduce extra leaks.
Rules \textsc{t-local} and \textsc{t-com} type the two constructs that \emph{write} into a store location, respectively a local assignment and a communication.
These are the only rules that enforce an actual security condition:
intuitively, the target must be at least as confidential as both the data being written and the control context in which the write happens.
Note that rule \textsc{t-com} treats communication as assignment between processes, thus assuming communication channels are private.
Rule \textsc{t-sel} types selections: since it only sends a fixed label $L$ it introduces no explicit flow and any implicit flow is already captured by the current $pc$.
Rule \textsc{t-cond} accounts for \emph{implicit flows}, it types a conditional by first typing the guard and then typing both branches under an updated program-counter label that tracks the sensitivity of the guard.
In other words, \textsc{t-cond} is the only rule that can raise $pc$; all other rules preserve it.
Rule \textsc{t-proc} types procedure calls by consulting the analysis context $\Delta$.  
The role of $\Delta$ is to summarise the \emph{admissible calling contexts} for each procedure as a function of $pc$: a call is accepted only if the caller’s typing environment matches one of the contexts recorded in $\Delta$ up to renaming of formal parameters.
This is the key mechanism that supports recursion without repeatedly re-checking bodies, while ensuring that procedures are only invoked under the security assumptions established for them.
To support this mechanism, we introduce new notation: $\bar{q}$ are formal parameters of procedure $X$, and $\bar{p}$ are arguments at the call site. Context renaming $\Gamma[\bar{q}\mapsto \bar{p}]$ updates $\Gamma$ so each $q_i$ points to the process security labelling of the respective $p_i$. Restricted equality $\Gamma \equiv_S \Gamma'$ means $\Gamma$ and $\Gamma'$ agree on all variables of all processes in $S$.
Rules \textsc{t-nil} and \textsc{t-seq} are purely structural: $\mathbf{0}$ is always typable and sequencing just composes typable choreographies.

\section{Soundness}
\label{sec:soundness}

We now proceed to prove the soundness of our type system w.r.t.\@
termination-insensitive non-interference. To specify the theorem, we introduce some notation.

\paragraph{Natural Semantics.}
We introduce a natural semantics~\cite{kahn1987natural}. We define relation
\[
\langle C, \Sigma, \mathscr{C} \rangle \Downarrow^M \Sigma'
\]
where $M$ is a sequence of transition labels, defined as the smallest relation:
\begin{mathpar}
\inferrule{}{\langle \boldsymbol{0}, \Sigma, \mathscr{C} \rangle \Downarrow^{[]} \Sigma}
\and
\inferrule{\langle C, \Sigma, \mathscr{C} \rangle \xrightarrow{\mu} \langle C', \Sigma', \mathscr{C} \rangle \\ \langle C', \Sigma', \mathscr{C} \rangle \Downarrow^M \Sigma''}{\langle C, \Sigma, \mathscr{C} \rangle \Downarrow^{\mu :: M} \Sigma''}
\end{mathpar}

\paragraph{Attacker Observation Model.}
We formalise $low$-level observation by defining a $low$-equivalence relation
$\Sigma_1 \equiv^\Gamma_{low} \Sigma_2$, which holds when the two stores agree
on all variables whose labels in $\Gamma$ are $\sqsubseteq low$. The only
$low$-observable outputs are the values of such $low$-security variables.

\bigskip
\noindent Then, we state key assumptions needed for soundness:

\paragraph{Well-formed Procedure Context.}
For every definition $X(\bar{p}) = C \in \mathscr{C}$:
\begin{equation}\label{ass:wellf_ctx}
\pn(C) \subseteq \{\bar{p}\}
\end{equation}

\paragraph{Well-typed Security Procedure Context.}
For every $X(\bar{p}) = C \in \mathscr{C}$ and $pc \in \MCL$:
\begin{equation}\label{ass:wellt_ctx}
\Delta\,X\,pc = \mathcal{G} \;\Longrightarrow\; \forall\,\Gamma \in \mathcal{G}.\, \Delta;\Gamma;pc \vdash C
\end{equation}
where \(\mathcal{G}\) is the set of admissible environments for \(C\) at program counter \(pc\).

\paragraph{Deterministic Expressions.}
Given process store $\sigma$, expression
$e$, and values $v_1, v_2$:
\[
\sigma \vdash e \downarrow v_1 \Rightarrow \sigma \vdash e \downarrow v_2 \Rightarrow v_1 = v_2
\]

\subsection{Instrumented Choreographies}
We prove soundness by induction on the length of computations. Since
non-interference of a configuration may depend on its execution history, we must
carry information across execution steps. To this end, we introduce
\emph{Instrumented Choreographies} as an extension of Recursive
Choreographies. In this variant, explicit \emph{brackets} delimit portions of code
that are allowed to differ across alternative executions. Beyond brackets, there
are two further differences w.r.t.\@ Recursive Choreographies:
\begin{itemize}
	\item \emph{Flattening of instructions and choreographies:} Brackets may
	enclose sequences of instructions of arbitrary length (including empty ones),
	making terms such as $[\boldsymbol{0}]$ and $\boldsymbol{0}$ distinct.
	\item \emph{Removal of the runtime term and label selection instructions:}
	Although not essential, this simplification eases the soundness proof. Its
	rationale is that both instructions have no impact on the final computed stores
	and are only needed in the semantics to support
	projection~\cite{montesi2023introduction}, which is omitted here.
\end{itemize}

\noindent The extended syntax supports a weaker notion of equivalence than syntactic
equality. We write $C_1 \approx_{low} C_2$ for \emph{$low$-equivalence} between
choreographies: the relation compares choreographies structurally while ignoring
the contents of bracketed subterms, equating them regardless of their internal
structure.

\paragraph{Instrumented stores.}
We define $\mathsf{[CStore]}$, an extension of $\mathsf{CStore}$ containing possibly bracketed values $\mathsf{[Val]}$. Formally:
\[
\mathsf{[Val]} ::= v \mid [v] \quad\text{with } v \in \mathsf{Val}
\]
To relate instrumented and reference stores, we introduce two notions:
\emph{well-formedness} and \emph{correctness}.

\paragraph{} A $\mathsf{[CStore]}$ $[\Sigma]$ is \emph{well formed} w.r.t.\ $\Gamma$ when for all $p, x, v$:
\[
[\Sigma]\,p.x = [v] \iff \Gamma\,p.x \not\sqsubseteq low
\]
Where $\not\sqsubseteq$ encodes observability: $a \not\sqsubseteq low$ means $low$ cannot see the value of $a$.

\paragraph{} We define a \emph{lowering} function $\lfloor [\Sigma] \rfloor$ that removes
brackets from values while leaving unbracketed values unchanged.
An instrumented store $[\Sigma]$ is \emph{correct} with respect to a reference
store $\Sigma$ when $\lfloor [\Sigma] \rfloor = \Sigma$.

\paragraph{Low-equivalence of configurations.}
The notion of $low$-equivalence extends naturally to instrumented stores:
two stores are $low$-equivalent when they coincide pointwise on all locations
after ignoring bracketed values.
We lift this notion to configurations: two configurations
$\langle [C_1], [\Sigma_1], \mathscr{C}\rangle$ and
$\langle [C_2], [\Sigma_2], \mathscr{C}\rangle$
are $low$-equivalent when $[C_1] \approx_{low} [C_2]$ and
$[\Sigma_1] \approx_{low} [\Sigma_2]$.

\paragraph{Instrumented semantics.}
The instrumented semantics extends the reference semantics with explicit
handling of bracketed values and choreographies and is parametric in the
flow-policy $\Gamma$ and the observation level $low$. Its purpose is twofold:
to maintain well-formedness of instrumented stores throughout execution, and to
record control-flow dependencies on high information by introducing brackets
around branching continuations when needed.

\paragraph{Type System Extension.}
We add a rule to the type system so that a bracketed choreography can be typed
at a lower program counter as long as its body executes at a strictly high
security level, isolating high effects within brackets.

\paragraph{Relating Reference and Instrumented Executions.}
To relate the reference semantics to the instrumented one, it is helpful to
think of the latter as a \emph{decorated replay} of an ordinary execution.
A \emph{lifting} of a reference execution is then an instrumented execution
that follows the same computational choices, but additionally records where
high data/control may have influenced the run by introducing brackets around
the corresponding continuations and values.

\subsection{Main Theorem}

\begin{theorem}[Termination-Insensitive Non-Interference]\label{thm:main}
Let $C$ be a choreography that is well-typed under the judgement $\Delta;\Gamma;\bot \vdash C$.
Given two choreographic stores $\Sigma_1, \Sigma_2$,
satisfying the $low$-equivalence relation $\Sigma_1 \equiv^\Gamma_{low} \Sigma_2$, suppose there exist terminating executions: 
$$\langle C, \Sigma_1, \mathscr{C}\rangle \Downarrow^{M_1} \Sigma'_1 \quad \text{and}\quad \langle C, \Sigma_2, \mathscr{C}\rangle \Downarrow^{M_2} \Sigma'_2$$
then the resulting stores satisfy the relation $\Sigma'_1 \equiv^\Gamma_{low} \Sigma'_2$.
\end{theorem}

\paragraph{Proof sketch.}
Conceptually, the proof proceeds by constructing corresponding executions in the
instrumented semantics and then applying a standard progress and preservation
style argument~\cite{pierce2002types}, where $low$-equivalence is shown to be
invariant along instrumented reductions. We now present the main lemmas
necessary to complete the argument:
\begin{itemize}
\item \emph{Completeness}: this lemma establishes a correspondence between the
	reference and the instrumented semantics. In particular, each terminating
	execution can be \emph{lifted} to an instrumented execution whose resulting
	stores are \emph{correct} w.r.t.\@ $\Sigma'_1, \Sigma'_2$, and
	\emph{well-formed} w.r.t.\@ $\Gamma$.
\item \emph{Preservation}: the typing judgement is preserved by the semantics,
i.e., starting from a well-typed configuration, each reduction step produces a
well-typed configuration. Here, a well-typed configuration consists of a
\emph{well-typed choreography} and a \emph{well-formed store}. Handling mutually
recursive procedures relies crucially on the \emph{well-typed security procedure
context} assumption.
\item \emph{Unwinding}: under the appropriate typing and well-formedness
assumptions, the instrumented semantics preserves $low$-equivalence of
configurations at each reduction step. This lemma motivates the instrumented
syntax and semantics so that, while strict syntactic equality may fail to hold
stepwise, $low$-equivalence of instrumented choreographies is preserved.
\end{itemize}

\noindent The proof proceeds by induction on the number of reduction steps in the
instrumented executions, which we construct using the \emph{Completeness} lemma.
The zero length case follows clearly from the given definition of \emph{Natural
Semantics}. The $n + 1$ case follows by combining the \emph{Preservation} lemma
(which let us use the induction hypothesis) and the \emph{Unwinding} lemma
(which morally proves non-interference step by step).

Note that we only consider pairs of terminating executions; divergence is
intentionally ignored, in line with termination-insensitive non-interference.

\paragraph{} Full definitions and proofs are available in~\cite[Chapter 4]{B25}.

\section{Context Reconstruction}
\label{sec:delta}

We present an algorithm that reconstructs $\Delta$ from procedure definitions $\mathscr{C}$ through constraint generation and solving.

\paragraph{Constraint Generation Pass.}
A \emph{constraint} has the form $\Uppsi \sqsubseteq p.x$, where $\Uppsi$ is a
symbolic \emph{bound} built from located variables, a distinguished fresh
variable $\eta$, and lattice
operations. We write $\llbracket \Uppsi \rrbracket\,\Gamma\,pc$ for the
evaluation of $\Uppsi$ under an environment $\Gamma$ and a program-counter label
$pc$. The evaluation of a constraint results in a truth value.

The constraint-generation algorithm $\delta,\eta \vdash C \rhd E$ is defined by
recursion on choreographies: it returns a finite set of constraints $E$, given a
constraint context $\delta$ (mapping procedure names to constraint sets) and a
fresh variable $\eta$ used to represent bounds on $pc$. For each procedure
$X(\bar p)=C$ we traverse $C$ and emit exactly the ordering requirements
suggested by the typing rules from Section~\ref{sec:types}, while keeping
expression levels symbolic. For instance, $p.x := e$ generates constraint
$\Uppsi_e \sqcup \eta \sqsubseteq p.x$ where $\Uppsi_e$ symbolically represents
$e$'s security level and $\eta$ represents the security level of $pc$. The main
difference with the typing rules is the treatment of procedure calls: a call
imports the constraint set already associated with the callee in $\delta$,
renaming formal parameters to actual arguments. This \emph{constraint forwarding} is
also the key intuition behind monotonicity in $\delta$ of the algorithm.

\paragraph{Iteration.}
A single reconstruction pass produces constraints for one procedure body under a given $\delta$.
We define an operator $\upphi_\mathscr{C}$ that maps $\delta$ to a new context $\delta'$ by running
constraint generation once for each procedure body in $\mathscr{C}$ and collecting the resulting constraint sets.
Since $\upphi_\mathscr{C}$ is monotonic, Knaster-Tarski's theorem~\cite{tarski1955lattice}
ensures the existence of a \emph{least} fixed point, denoted $\mu \upphi_\mathscr{C}$.

\paragraph{Generating $\Delta$.}
We define $ \Delta \triangleq \underline{gen}~(\mu\upphi_\mathscr{C}) $
where $\underline{gen}~\delta$ creates a map from procedure and $pc$ to the set of environments that satisfy constraints in $\delta$, formally:
\[
\Delta~(X,pc) \triangleq
\left\{ \Gamma \ \middle|\ 
\forall (\Uppsi \sqsubseteq p.x) \in E_X,\ 
\llbracket \Uppsi \rrbracket\,\Gamma\,pc \sqsubseteq \Gamma~p.x
\right\},
\]
here $E_X$ is the constraint set associated with $X$ in $\mu\upphi_\mathscr{C}$.

\subsection{Correctness of Reconstruction}
We now show that the generated context is correct with respect to the typing relation.

\begin{theorem}[Well-typedness of the generated procedure context]
\label{thm:gen-welltyped}
Let $\Delta \triangleq \underline{gen}~(\mu \upphi_{\mathscr C})$.
Then, for every procedure identifier $X$, program-counter label $pc$, and environment $\Gamma$,
if $\Gamma \in \Delta~X~pc$, it holds that $\Delta,\Gamma,pc \vdash C_X$.
\end{theorem}

\paragraph{Proof sketch.}
The key lemma is a one-step soundness property: if reconstruction produces constraints $E$ for a choreography $C$ under some $\delta$ and an environment $\Gamma$ satisfies $E$, then $C$ is typable under $\Delta=\underline{gen}~\delta$ and $\Gamma$.
Intuitively, constraint reconstruction is just the typing rules \emph{run backwards}: for each construct it emits exactly the flow-inequalities needed as rule premises, so any $\Gamma$ that satisfies the generated constraints immediately provides the hypotheses to rebuild the typing derivation for $C$.
Finally, let $\delta^\star = \mu\upphi_{\mathscr C}$; since $\delta^\star$ is a fixed point, for every procedure $X$ the constraint set stored at $\delta^\star~X$ is exactly the one reconstructed from $C_X$ under $\delta^\star$, so any $\Delta^\star = \underline{gen}~\delta^\star$, $\Gamma\in\Delta^\star~X~pc$ satisfies the hypotheses of one-step soundness and hence $\Delta^\star,\Gamma,pc\vdash C_X$.

\paragraph{Termination.}
Since all the functions in this section are defined over finite structures, there will always be a terminating algorithm to compute them.

\paragraph{} Full definitions and proofs are available in~\cite[Chapter 5]{B25}.

\section{Related Work}
\label{sec:related_work}

A closely related line of work is due to Lluch-Lafuente et al.~\cite{LLNN15}, who study information-flow control for choreographic programs. Their approach targets \emph{discretionary} information-flow policies, expressed as access-control constraints that regulate which principals may read or update particular data items. They instrument the operational semantics with rich information-flow annotations and develop a sound type system that statically checks compliance of a choreographic program with a given policy. 
In contrast, we establish the stronger semantic property of termination-insensitive non-interference.

More broadly, information-flow security has a long history, beginning with Denning's lattice model~\cite{denning1976lattice} and the seminal formulation of non-interference by Goguen and Meseguer~\cite{goguen1982security}. Volpano et
al.~\cite{volpano1996sound} introduced one of the first type systems for secure information-flow analysis in an imperative setting, and a substantial body of subsequent work has refined and generalised type-based enforcement mechanisms for sequential languages~\cite{sabelfeld2003language,myers2011proving}.
These systems provide static, compositional guarantees on both explicit and implicit flows, typically via a program-counter discipline closely related to the one we adopt. Our work builds directly on this line of research. The key difference is that we lift these foundational ideas from sequential, local programs to a distributed setting in which communication and control flow are specified globally as choreographies. 

Dynamic and hybrid techniques for information-flow security have also been
extensively studied. For instance, language-based runtime enforcement
mechanisms~\cite{sabelfeld2003language} can monitor executions and prevent
insecure behaviour at run time. More recent work on permissive runtime
information-flow control~\cite{BRGH21} improves precision in the presence of
language features such as exceptions. However, purely dynamic mechanisms
inherently reason about single concrete executions, whereas non-interference is
a relational property over multiple executions. As a result, dynamic enforcement
alone cannot generally establish semantic non-interference, and extending such
guarantees to globally specified distributed protocols can pose additional
challenges, especially for certain classes of implicit flows. In contrast, our
approach provides a static, compositional proof of termination-insensitive
non-interference for choreographic programs.

Our work also relates to type systems for concurrent calculi and session-based formalisms with security annotations. 
Bhargavan et al.~\cite{BCDFL09} and Capecchi et al.~\cite{CCDR10} were among the first to integrate access control and information-flow guarantees into session-typed process calculi, combining protocol structure with security levels. 
Subsequent work on secure session types and multiparty protocols establishes access-control or information-flow properties at the level of local endpoint processes~\cite{CCD16,CDP16}.
The distinction between these approaches and ours mirrors the broader difference between multiparty session types and choreographic programming. Session-based approaches typically verify endpoint implementations against a global type specification, deriving local types used to type-check processes whose internal computations are abstracted by the type.
In contrast, we verify a full distributed implementation written as a choreographic program, which specifies not only the communication structure but also the concrete data manipulations and control flow. 
Endpoint implementations are then obtained by projection. 
Consequently, non-interference is established once at the global program level and inherited by the projected endpoints by construction.

Security protocol verification tools such as ProVerif~\cite{B16} and
Tamarin~\cite{MSCB13} provide powerful automated support for checking
cryptographic protocols in symbolic models. These approaches excel at modelling
cryptographic primitives and attacker capabilities, and can verify sophisticated
authentication and secrecy properties. However, they operate over specialised 
protocol formalisms and establish properties by model checking specific 
protocol instances against user-specified security goals. They do not aim to 
provide a type-based, compositional characterisation of non-interference 
integrated into a programming model for distributed applications. 
Our approach is complementary. We equip the reference choreographic programming language with a static type system that supports compositional, static type checking and reconstruction, and we prove a general syntactic non-interference theorem. Rather than analysing protocols instance by instance, we establish a language-level guarantee that applies uniformly to all well-typed choreographic programs.

Choreographic programming has been extensively developed as a foundation for correct-by-construction distributed programming~\cite{montesi2023introduction}. Its central result is endpoint projection: a choreographic program can be compiled into local endpoint implementations that are correct by construction, guaranteeing properties such as deadlock-freedom and communication safety~\cite{M13,CM20,BLT20,JB22,BLT23,GMP25,PPM24}.
Foundational work established the semantic correctness of projection and its connections to multiparty session types and global types~\cite{CDP12,CM13,DY12} and have supported further developments such as model-based synthesis~\cite{ARSIT13}, logics for reasoning about choreographic programs \cite{CMS18,CGMP23}, and mechanically certified compilation to endpoint code~\cite{CMP21,CMP23,PGSN22,HG22}.
These theoretical foundations have been realised with the implementation of several choreographic programming languages, including Chor~\cite{CM13}, AIOCJ~\cite{DGGLM17} and, more recently, Choral~\cite{GMP24}, and HasChor~\cite{SKK23}, which integrate choreographic programming with mainstream
object-oriented and functional programming languages and explore advanced
features such as higher-order choreographies, location polymorphism, and
library-level endpoint projection. 
Across this body of work, the primary focus has been functional correctness and communication safety whereas our work is the first to address non-interference.

Finally, our proof methodology builds on standard syntactic techniques for
establishing type soundness and security properties. We follow the
progress and preservation style of Pierce~\cite{pierce2002types} and Wright and
Felleisen~\cite{wright1994syntactic}, and we make use of natural semantics in
the style of Kahn~\cite{kahn1987natural} to formulate termination-insensitive
non-interference. The use of a bracketed, instrumented semantics is by now
standard in language-based information-flow
security~\cite{sabelfeld2003language,myers2011proving}.

\section{Conclusion}
\label{sec:conclusion}

We have presented a type-based technique for non-interference in choreographic
programs. Our type system combines a policy-parametric security lattice with a
program-counter discipline to prevent information leaks through both explicit
data flows and implicit control dependencies. Well-typed choreographies satisfy
termination-insensitive non-interference, ensuring that high-security data
cannot influence low-security observations. Our type system supports procedural
choreographies through a procedure context for admissible calling environments.
We formalised a constraint-based reconstruction algorithm that computes said
context, eliminating the need for manual security annotations.

We see several directions for future work. First, we plan to extend our type
system to handle more expressive security policies, including declassification
mechanisms that permit controlled information release under specified
conditions. Second, we intend to explore other security properties, moving
beyond termination-insensitive non-interference. Indeed, \emph{side
channels}~\cite{kelsey1998side} (e.g., timing, termination, resource usage,
cache effects, message sizes, scheduler-dependent behaviour) fall outside this
view, unless the semantics and the attacker observation model explicitly make
them observable. Our termination-insensitive approach deliberately abstracts
from timing and divergence, which suffices for many practical scenarios but does
not address all potential covert channels. When side channels matter, one can
bring them into scope by enriching the semantics with cost or timing observables
and adopting timing-/step-sensitive definitions~\cite{197207}. Finally, we plan
to mechanise the entire development in a proof assistant such as Lean or Rocq.
While the pen-and-paper proofs provide confidence in our results, a complete
mechanisation would offer machine-checked assurance and eliminate the
possibility of subtle errors in lengthy case analyses and inductive arguments.
Besides strengthening confidence in our results, successfully completing this
mechanisation would provide a foundation for developing certified type checkers
and verified compilation tool-chains for secure choreographic programming. We
envision these checkers as bridges to bring our theoretical results to practical
implementation, e.g., by integrating them within existing choreographic
programming frameworks, such as Choral~\cite{GMP24}.

\bibliographystyle{splncs04}
\bibliography{refs}

@inproceedings{LLNN15,
  author       = {Alberto Lluch{-}Lafuente and
                  Flemming Nielson and
                  Hanne Riis Nielson},
  editor       = {Narciso Mart{\'{\i}}{-}Oliet and
                  Peter Csaba {\"{O}}lveczky and
                  Carolyn L. Talcott},
  title        = {Discretionary Information Flow Control for Interaction-Oriented Specifications},
  booktitle    = {Logic, Rewriting, and Concurrency - Essays dedicated to Jos{\'{e}}
                  Meseguer on the Occasion of His 65th Birthday},
  series       = {Lecture Notes in Computer Science},
  volume       = {9200},
  pages        = {427--450},
  publisher    = {Springer},
  year         = {2015},
  doi          = {10.1007/978-3-319-23165-5\_20},
}

@masterthesis{B25,
  title         = {Mechanized Type-Based Enforcement of Non-Interference in Choreographic Languages},
  author        = {Bertoni, Marco},
  url           = {https://amslaurea.unibo.it/id/eprint/36727/}
}

@book{montesi2023introduction,
  title         = {Introduction to Choreographies},
  author        = {Montesi, Fabrizio},
  year          = {2023},
  publisher     = {Cambridge University Press}
}

@article{plotkin2004origins,
  title         = {The origins of structural operational semantics},
  author        = {Plotkin, Gordon D},
  year          = {2004},
  journal       = {The Journal of Logic and Algebraic Programming},
  publisher     = {Elsevier},
  volume        = {60},
  pages         = {3--15}
}

@inproceedings{kahn1987natural,
  title         = {Natural semantics},
  author        = {Kahn, Gilles},
  year          = {1987},
  booktitle     = {Annual symposium on theoretical aspects of computer science},
  pages         = {22--39},
  organization  = {Springer}
}

@article{denning1976lattice,
  title         = {A lattice model of secure information flow},
  author        = {Denning, Dorothy E},
  year          = {1976},
  journal       = {Communications of the ACM},
  publisher     = {ACM New York, NY, USA},
  volume        = {19},
  number        = {5},
  pages         = {236--243}
}

@inproceedings{goguen1982security,
  title         = {Security policies and security models},
  author        = {Goguen, Joseph A and Meseguer, Jos{\'e}},
  year          = {1982},
  booktitle     = {1982 IEEE Symposium on Security and Privacy},
  pages         = {11--11},
  organization  = {IEEE}
}

@article{volpano1996sound,
  title         = {A sound type system for secure flow analysis},
  author        = {Volpano, Dennis and Irvine, Cynthia and Smith, Geoffrey},
  year          = {1996},
  journal       = {Journal of computer security},
  publisher     = {SAGE Publications Sage UK: London, England},
  volume        = {4},
  number        = {2-3},
  pages         = {167--187}
}

@article{sabelfeld2003language,
  title         = {Language-based information-flow security},
  author        = {Sabelfeld, Andrei and Myers, Andrew C},
  year          = {2003},
  journal       = {IEEE Journal on selected areas in communications},
  publisher     = {IEEE},
  volume        = {21},
  number        = {1},
  pages         = {5--19}
}

@article{myers2011proving,
  title         = {Proving noninterference for a while-language using small-step operational semantics},
  author        = {Myers, Andrew},
  year          = {2011}
}

@article{wright1994syntactic,
  title         = {A syntactic approach to type soundness},
  author        = {Wright, Andrew K and Felleisen, Matthias},
  year          = {1994},
  journal       = {Information and computation},
  publisher     = {Elsevier},
  volume        = {115},
  number        = {1},
  pages         = {38--94}
}

@inproceedings{197207,
  title         = {Verifying {Constant-Time} Implementations},
  author        = {Jose Bacelar Almeida and Manuel Barbosa and Gilles Barthe and Fran{\c c}ois Dupressoir and Michael Emmi},
  year          = {2016},
  month         = aug,
  booktitle     = {25th USENIX Security Symposium (USENIX Security 16)},
  publisher     = {USENIX Association},
  address       = {Austin, TX},
  pages         = {53--70},
  isbn          = {978-1-931971-32-4},
  url           = {https://www.usenix.org/conference/usenixsecurity16/technical-sessions/presentation/almeida}
}

@inproceedings{kelsey1998side,
  title         = {Side channel cryptanalysis of product ciphers},
  author        = {Kelsey, John and Schneier, Bruce and Wagner, David and Hall, Chris},
  year          = {1998},
  booktitle     = {European Symposium on Research in Computer Security},
  pages         = {97--110},
  organization  = {Springer}
}

@book{pierce2002types,
  title         = {Types and programming languages},
  author        = {Pierce, Benjamin C},
  year          = {2002},
  publisher     = {MIT press}
}

@article{tarski1955lattice,
  title         = {A lattice-theoretical fixpoint theorem and its applications.},
  author        = {Tarski, Alfred},
  year          = {1955}
}

@article{BRGH21,
  title         = {Permissive runtime information flow control in the presence of exceptions},
  author        = {Abhishek Bichhawat and Vineet Rajani and Deepak Garg and Christian Hammer},
  year          = {2021},
  journal       = {J. Comput. Secur.},
  volume        = {29},
  number        = {4},
  pages         = {361--401},
  doi           = {10.3233/JCS-211385}
}

@inproceedings{CCDR10,
  title         = {Session Types for Access and Information Flow Control},
  author        = {Sara Capecchi and Ilaria Castellani and Mariangiola Dezani{-}Ciancaglini and Tamara Rezk},
  year          = {2010},
  booktitle     = {{CONCUR} 2010 - Concurrency Theory, 21th International Conference, {CONCUR} 2010, Paris, France, August 31-September 3, 2010. Proceedings},
  publisher     = {Springer},
  series        = {Lecture Notes in Computer Science},
  volume        = {6269},
  pages         = {237--252},
  doi           = {10.1007/978-3-642-15375-4\_17},
  editor        = {Paul Gastin and Fran{\c{c}}ois Laroussinie}
}

@article{B16,
  title         = {Modeling and Verifying Security Protocols with the Applied Pi Calculus and ProVerif},
  author        = {Bruno Blanchet},
  year          = {2016},
  journal       = {Found. Trends Priv. Secur.},
  volume        = {1},
  number        = {1-2},
  pages         = {1--135},
  doi           = {10.1561/3300000004}
}

@inproceedings{MSCB13,
  title         = {The {TAMARIN} Prover for the Symbolic Analysis of Security Protocols},
  author        = {Simon Meier and Benedikt Schmidt and Cas Cremers and David A. Basin},
  year          = {2013},
  booktitle     = {Computer Aided Verification - 25th International Conference, {CAV} 2013, Saint Petersburg, Russia, July 13-19, 2013. Proceedings},
  publisher     = {Springer},
  series        = {Lecture Notes in Computer Science},
  volume        = {8044},
  pages         = {696--701},
  doi           = {10.1007/978-3-642-39799-8\_48},
  editor        = {Natasha Sharygina and Helmut Veith}
}

@phdthesis{M13,
  title         = "Choreographic {P}rogramming",
  author        = "Fabrizio Montesi",
  year          = 2013,
  note          = {\url{https://www.fabriziomontesi.com/files/choreographic-programming.pdf}},
  school        = "IT University of Copenhagen",
  type          = "Ph.{D}. Thesis"
}

@article{CM20,
  title         = {A core model for choreographic programming},
  author        = {Lu{\'{\i}}s Cruz{-}Filipe and Fabrizio Montesi},
  year          = {2020},
  journal       = {Theor. Comput. Sci.},
  volume        = {802},
  pages         = {38--66},
  doi           = {10.1016/J.TCS.2019.07.005}
}

@article{CMP23,
  title         = {A Formal Theory of Choreographic Programming},
  author        = {Lu{\'{\i}}s Cruz{-}Filipe and Fabrizio Montesi and Marco Peressotti},
  year          = {2023},
  journal       = {J. Autom. Reason.},
  volume        = {67},
  number        = {2},
  pages         = {21},
  doi           = {10.1007/S10817-023-09665-3}
}

@article{HG22,
  author       = {Andrew K. Hirsch and
                  Deepak Garg},
  title        = {Pirouette: higher-order typed functional choreographies},
  journal      = {Proc. {ACM} Program. Lang.},
  volume       = {6},
  number       = {{POPL}},
  pages        = {1--27},
  year         = {2022},
  url          = {https://doi.org/10.1145/3498684},
  doi          = {10.1145/3498684},
  bibsource    = {dblp computer science bibliography, https://dblp.org}
}

@inproceedings{PGSN22,
  author       = {Johannes {\AA}man Pohjola and
                  Alejandro G{\'{o}}mez{-}Londo{\~{n}}o and
                  James Shaker and
                  Michael Norrish},
  editor       = {June Andronick and
                  Leonardo de Moura},
  title        = {Kalas: {A} Verified, End-To-End Compiler for a Choreographic Language},
  booktitle    = {13th International Conference on Interactive Theorem Proving, {ITP}
                  2022, Haifa, Israel, August 7-10, 2022},
  series       = {LIPIcs},
  volume       = {237},
  pages        = {27:1--27:18},
  publisher    = {Schloss Dagstuhl - Leibniz-Zentrum f{\"{u}}r Informatik},
  year         = {2022},
  url          = {https://doi.org/10.4230/LIPIcs.ITP.2022.27},
  doi          = {10.4230/LIPICS.ITP.2022.27},
  bibsource    = {dblp computer science bibliography, https://dblp.org}
}

@article{CDP12,
  title         = {On Global Types and Multi-Party Session},
  author        = {Giuseppe Castagna and Mariangiola Dezani{-}Ciancaglini and Luca Padovani},
  year          = {2012},
  journal       = {Log. Methods Comput. Sci.},
  volume        = {8},
  number        = {1},
  doi           = {10.2168/LMCS-8(1:24)2012}
}

@inproceedings{CM13,
  title         = {Deadlock-freedom-by-design: multiparty asynchronous global programming},
  author        = {Marco Carbone and Fabrizio Montesi},
  year          = {2013},
  booktitle     = {The 40th Annual {ACM} {SIGPLAN-SIGACT} Symposium on Principles of Programming Languages, {POPL} '13, Rome, Italy - January 23 - 25, 2013},
  publisher     = {{ACM}},
  pages         = {263--274},
  doi           = {10.1145/2429069.2429101},
  editor        = {Roberto Giacobazzi and Radhia Cousot}
}

@inproceedings{DY12,
  title         = {Multiparty Session Types Meet Communicating Automata},
  author        = {Pierre{-}Malo Deni{\'{e}}lou and Nobuko Yoshida},
  year          = {2012},
  booktitle     = {Programming Languages and Systems - 21st European Symposium on Programming, {ESOP} 2012, Held as Part of the European Joint Conferences on Theory and Practice of Software, {ETAPS} 2012, Tallinn, Estonia, March 24 - April 1, 2012. Proceedings},
  publisher     = {Springer},
  series        = {Lecture Notes in Computer Science},
  volume        = {7211},
  pages         = {194--213},
  doi           = {10.1007/978-3-642-28869-2\_10},
  editor        = {Helmut Seidl}
}

@article{DGGLM17,
  title         = {Dynamic Choreographies: Theory And Implementation},
  author        = {Mila {Dalla Preda} and Maurizio Gabbrielli and Saverio Giallorenzo and Ivan Lanese and Jacopo Mauro},
  year          = {2017},
  journal       = {Log. Methods Comput. Sci.},
  volume        = {13},
  number        = {2},
  doi           = {10.23638/LMCS-13(2:1)2017}
}

@article{GMP24,
  title         = {Choral: Object-oriented Choreographic Programming},
  author        = {Saverio Giallorenzo and Fabrizio Montesi and Marco Peressotti},
  year          = {2024},
  journal       = {{ACM} Trans. Program. Lang. Syst.},
  volume        = {46},
  number        = {1},
  pages         = {1:1--1:59},
  doi           = {10.1145/3632398}
}

@article{SKK23,
  title         = {HasChor: Functional Choreographic Programming for All (Functional Pearl)},
  author        = {Gan Shen and Shun Kashiwa and Lindsey Kuper},
  year          = {2023},
  journal       = {Proc. {ACM} Program. Lang.},
  volume        = {7},
  number        = {{ICFP}},
  pages         = {541--565},
  doi           = {10.1145/3607849}
}

@inproceedings{ARSIT13,
  title         = {A Model-Based Synthesis Process for Choreography Realizability Enforcement},
  author        = {Marco Autili and Davide Di Ruscio and Amleto Di Salle and Paola Inverardi and Massimo Tivoli},
  year          = {2013},
  booktitle     = {Fundamental Approaches to Software Engineering - 16th International Conference, {FASE} 2013, Held as Part of the European Joint Conferences on Theory and Practice of Software, {ETAPS} 2013, Rome, Italy, March 16-24, 2013. Proceedings},
  publisher     = {Springer},
  series        = {Lecture Notes in Computer Science},
  volume        = {7793},
  pages         = {37--52},
  doi           = {10.1007/978-3-642-37057-1\_4},
  editor        = {Vittorio Cortellessa and D{\'{a}}niel Varr{\'{o}}}
}

@inproceedings{BLT20,
  title         = {Choreography Automata},
  author        = {Franco Barbanera and Ivan Lanese and Emilio Tuosto},
  year          = {2020},
  booktitle     = {Coordination Models and Languages - 22nd {IFIP} {WG} 6.1 International Conference, {COORDINATION} 2020, Held as Part of the 15th International Federated Conference on Distributed Computing Techniques, DisCoTec 2020, Valletta, Malta, June 15-19, 2020, Proceedings},
  publisher     = {Springer},
  series        = {Lecture Notes in Computer Science},
  volume        = {12134},
  pages         = {86--106},
  doi           = {10.1007/978-3-030-50029-0\_6},
  editor        = {Simon Bliudze and Laura Bocchi}
}

@inproceedings{JB22,
  title         = {A Predicate Transformer for Choreographies - Computing Preconditions in Choreographic Programming},
  author        = {Sung{-}Shik Jongmans and Petra van den Bos},
  year          = {2022},
  booktitle     = {Programming Languages and Systems - 31st European Symposium on Programming, {ESOP} 2022, Held as Part of the European Joint Conferences on Theory and Practice of Software, {ETAPS} 2022, Munich, Germany, April 2-7, 2022, Proceedings},
  publisher     = {Springer},
  series        = {Lecture Notes in Computer Science},
  volume        = {13240},
  pages         = {520--547},
  doi           = {10.1007/978-3-030-99336-8\_19},
  editor        = {Ilya Sergey}
}

@article{CMS18,
  author       = {Marco Carbone and
                  Fabrizio Montesi and
                  Carsten Sch{\"{u}}rmann},
  title        = {Choreographies, logically},
  journal      = {Distributed Comput.},
  volume       = {31},
  number       = {1},
  pages        = {51--67},
  year         = {2018},
  url          = {https://doi.org/10.1007/s00446-017-0295-1},
  doi          = {10.1007/S00446-017-0295-1},
  bibsource    = {dblp computer science bibliography, https://dblp.org}
}

@inproceedings{CGMP23,
  author = {Cruz{-}Filipe, Lu{\'{\i}}s and Graversen, Eva and Montesi, Fabrizio and Peressotti, Marco},
  editor = {Jongmans, Sung{-}Shik and Lopes, Ant{\'{o}}nia},
  title = {Reasoning About Choreographic Programs},
  booktitle = {Coordination Models and Languages},
  series = {Lecture Notes in Computer Science},
  volume = {13908},
  pages = {144--162},
  publisher = {Springer},
  year = {2023},
  doi = {10.1007/978-3-031-35361-1\_8},
  isbn = {978-3-031-35361-1},
  wos = {WOS:001288400900008},
  scopus = {2-s2.0-85164736949}
}

@inproceedings{GMP25,
  author = {Graversen, Eva and Montesi, Fabrizio and Peressotti, Marco},
  editor = {Moscardelli, Luca},
  title = {Omission Failures in Choreographic Programming},
  booktitle = {Proceedings of the 26th Italian Conference on Theoretical Computer Science, Pescara, Italy, September 10-12, 2025},
  series = {{CEUR} Workshop Proceedings},
  volume = {4039},
  pages = {169--184},
  publisher = {CEUR-WS.org},
  year = {2025},
  url = {https://ceur-ws.org/Vol-4039/paper02.pdf}
}

@inproceedings{PPM24,
  author = {Plyukhin, Dan and Peressotti, Marco and Montesi, Fabrizio},
  editor = {Aldrich, Jonathan and Salvaneschi, Guido},
  title = {Ozone: Fully Out-of-Order Choreographies},
  booktitle = {38th European Conference on Object-Oriented Programming, {ECOOP} 2024, September 16-20, 2024, Vienna, Austria},
  series = {LIPIcs},
  volume = {313},
  pages = {31:1--31:28},
  publisher = {Schloss Dagstuhl - Leibniz-Zentrum f{\"{u}}r Informatik},
  year = {2024},
  doi = {10.4230/LIPICS.ECOOP.2024.31},
  scopus = {2-s2.0-85204986670}
}

@article{BLT23,
  title         = {A Theory of Formal Choreographic Languages},
  author        = {Franco Barbanera and Ivan Lanese and Emilio Tuosto},
  year          = {2023},
  journal       = {Log. Methods Comput. Sci.},
  volume        = {19},
  number        = {3},
  doi           = {10.46298/LMCS-19(3:9)2023}
}

@inproceedings{CMP21,
  title         = {Certifying Choreography Compilation},
  author        = {Lu{\'{\i}}s Cruz{-}Filipe and Fabrizio Montesi and Marco Peressotti},
  year          = {2021},
  booktitle     = {Theoretical Aspects of Computing - {ICTAC} 2021 - 18th International Colloquium, Virtual Event, Nur-Sultan, Kazakhstan, September 8-10, 2021, Proceedings},
  publisher     = {Springer},
  series        = {Lecture Notes in Computer Science},
  volume        = {12819},
  pages         = {115--133},
  doi           = {10.1007/978-3-030-85315-0\_8},
  editor        = {Antonio Cerone and Peter Csaba {\"{O}}lveczky}
}

@article{CCD16,
  title         = {Information flow safety in multiparty sessions},
  author        = {Sara Capecchi and Ilaria Castellani and Mariangiola Dezani{-}Ciancaglini},
  year          = {2016},
  journal       = {Math. Struct. Comput. Sci.},
  volume        = {26},
  number        = {8},
  pages         = {1352--1394},
  doi           = {10.1017/S0960129514000619}
}

@article{CDP16,
  title         = {Self-adaptation and secure information flow in multiparty communications},
  author        = {Ilaria Castellani and Mariangiola Dezani{-}Ciancaglini and Jorge A. P{\'{e}}rez},
  year          = {2016},
  journal       = {Formal Aspects Comput.},
  volume        = {28},
  number        = {4},
  pages         = {669--696},
  doi           = {10.1007/S00165-016-0381-3}
}

@inproceedings{BCDFL09,
  title         = {Cryptographic Protocol Synthesis and Verification for Multiparty Sessions},
  author        = {Karthikeyan Bhargavan and Ricardo Corin and Pierre{-}Malo Deni{\'{e}}lou and C{\'{e}}dric Fournet and James J. Leifer},
  year          = {2009},
  booktitle     = {Proceedings of the 22nd {IEEE} Computer Security Foundations Symposium, {CSF} 2009, Port Jefferson, New York, USA, July 8-10, 2009},
  publisher     = {{IEEE} Computer Society},
  pages         = {124--140},
  doi           = {10.1109/CSF.2009.26}
}

\end{document}